\documentclass[lettersize,journal,onecolumn]{IEEEtran}

\usepackage{cite}
\usepackage{amsmath,amssymb,amsfonts,amsthm,mathrsfs}
\usepackage{multirow}
\usepackage{bbm}
\usepackage{algorithmic}
\usepackage{caption}
\usepackage{graphicx}
\usepackage{tikz} 
\usetikzlibrary{automata, positioning, arrows}
\usepackage{textcomp}
\usepackage{xcolor}
\usepackage{manfnt}
\usepackage{mathtools}
\usepackage{fixmath}
\usepackage{pgfplots}
\pgfplotsset{compat=1.14}
\usepackage{pdfpages}

\usepackage{xcolor}
\usetikzlibrary{positioning,arrows}
\usepackage{balance} 

\def\BibTeX{{\rm B\kern-.05em{\sc i\kern-.025em b}\kern-.08em
    T\kern-.1667em\lower.7ex\hbox{E}\kern-.125emX}}
    
\newcommand{\N}{\mathbb{N}}

\newcommand{\Q}{\mathbb{Q}}

\newcommand{\sM}{\mathcal{M}}

\theoremstyle{plain}
\newtheorem{thm}{Theorem}

\theoremstyle{defn}
\newtheorem{defn}{Definition}

\theoremstyle{rem}
\newtheorem{rem}{Remark}

\tikzstyle{block} = [draw, rectangle, 
  minimum height=3em, minimum width=4em]

\begin{document}

\title{Finite Blocklength Performance of Capacity-achieving Codes in the Light of Complexity Theory}

\author{\IEEEauthorblockN{Holger Boche\IEEEauthorrefmark{1}, Andrea Grigorescu\IEEEauthorrefmark{1}, Rafael F. Schaefer\IEEEauthorrefmark{2}, H. Vincent Poor\IEEEauthorrefmark{3}}\\
\IEEEauthorblockA{\small 
\IEEEauthorrefmark{1}\textit{Technical University of Munich}\\
\IEEEauthorrefmark{2}\textit{Technische Universit\"at Dresden}\\
\IEEEauthorrefmark{3}\textit{Princeton University}\\
\texttt{boche@tum.de, andrea.grigorescu@tum.de,rafael.schaefer@tu-dresden.de, poor@princeton.edu}
}}

\maketitle

Since the work of \cite{polyanskiy2010channel} on the finite blocklength performance of capacity-achieving codes for discrete memoryless channels, many papers have developed further results for several practically relevant channels, see \cite{polyanskiy2011dispersion,polyanskiy2011feedback,yang2019wiretap}. However, the complexity of computing capacity-achieving codes has not been investigated until now. We study this question for one of the simplest of non-trivial Gaussian channels, i.e., the additive colored Gaussian noise (ACGN) channel. 

In this context, it is essential to have a well-defined concept of complexity. In particular, the ``parameters'' of the communication system should be of low complexity, meaning they should be easy to describe. We focus on a point-to-point ACGN channel. This system is fully characterized by the transmission power $P$ of the transmitter and the noise power spectral density $N$. Therefore, both the transmission power $P$ and the power spectral density $N$ should be easy to compute. The central question is how complex it is to compute key performance metrics of the communication system. Of particular practical importance are the Shannon capacity and sequences of capacity-achievable codes.

To assess computational complexity, we consider the classes $\mathrm{FP}$, $\mathrm{FP}_1$, and $\#\mathrm{P}_1$. For this, we first need to introduce the set of all finite strings over the binary alphabet, denoted by $\{0,1\}^*$.
\begin{defn}[Class $\mathrm{FP}$]\label{def:FPsharpP}
A function $f \colon \{0, 1\}^* \rightarrow \N$ is in $\mathrm{FP}$ if it can be computed by a deterministic Turing machine in polynomial time.
\end{defn}

In this work, we are also interested in studying functions that are defined on the singleton alphabet, i.e., $\{0\}^* \subset \{0, 1\}^*$. In other words, these functions are defined solely on the set of finite words composed of the symbol $0$. The class analog to $\mathrm{FP}$ defined on singleton sets are denoted by $\mathrm{FP}_1$. We further introduce the class $\#\mathrm{P}_1$ also defined on the singleton alphabet. $\#\mathrm{P}_1$ encompasses functions that count the number of solutions verifiable by a Turing machine in polynomial time.

\begin{defn}[Classes $\mathrm{FP}_1$ and $\#\mathrm{P}_1$]\label{def:FP1sharpP1}
A function $f \colon\{0\}^* \to \N$ is said to be in $\mathrm{FP}_1$ if it can be computed by a deterministic Turing machine in polynomial time.

A function $f\colon \{0\}^* \to \N$ is said to be in $\#\mathrm{P}_1$ if there exists a polynomial $p \colon \N \to \N$ and a polynomial
time Turing machine $M$ so that for every string $x \in \{0\}^*$
\begin{equation*}
	f(x)=|\{y\in\{0\}^{p(|x|)}\colon M(x,y)=1\}|.
\end{equation*} 
\end{defn}

In \cite{boche2024characterizationj}, it has been shown that there exists an infinitely differentiable, strictly positive noise power spectral density $N_*$, which is computable in polynomial time and such that for every sufficiently large rational power constraint $P$ under the widely accepted assumption $\mathrm{FP}_1 \neq \#\mathrm{P}_1$, the capacity $C(P, N_*)$ cannot be computed in polynomial time, demonstrating a complexity-blowup phenomenon.

\begin{figure}[h!]
\centering
\begin{tikzpicture}[scale=0.6]
	\draw [dotted] plot [domain=1:10,samples=100] (\x,{log2(\x)});
	\draw[-] [red] (0, 3.8) -- (10, 3.8) ;
	\node(a) at(0, 3.8) {};
  \draw[->] (0, 0) -- (10, 0) node[right] {$n$};
  \draw[->] (0, 0) -- (0, 4);
  \node [left of= a] {$C$};
  \node (C1) at (0,2.8) {-};
 \node (C2) at (0,1.8) {-};
  \node (C14) [ left of= C1] {$C-\frac{1}{4}$};
  \node (C12) [ left of= C2] {$C-\frac{1}{2}$};

    \node (n2) at (7.966, 0) {};
    \draw[-]  (7.966, -.1) --  (7.966, 0.1) ;
    \draw[loosely dotted] (0,2.92)-- (7.966, 2.92);
    \draw[loosely dotted] (7.966, 0)-- (7.966, 2.92);
    \draw[loosely dotted] (0,1.78)-- (3.482, 1.78);
    \draw[loosely dotted] (3.498, 0)-- (3.498, 1.76);
    \node (n1) at (3.482, 0) {};
    \draw[-]  (3.482, -.1) --  (3.482, 0.1) ;
  \node (C14) [ below of= n2] {$n_2$};
  \node (C12) [ below of= n1] {$n_1$};
  \node [black] at (7.966, 2.92) {\textbullet};
  \node [black] at (3.482, 1.76) {\textbullet};
\end{tikzpicture}\caption{The red line represents the band-limited ACGN capacity $C=C(P,N_*)$ for the power spectral density. $N_*$ and the power constraint $P$ in the asymptotic regime. The black curve represents the finite blocklength achievable rate $R_{n}(\epsilon)$ for some fixed $\epsilon>0$. For $n_1$ we have $R_{n_1}(\epsilon)>C-\frac{1}{2}$ and for $n_2$ we have $R_{n_2}(\epsilon)>C-\frac{1}{4}$. Fig. from \cite{boche2024characterizationj}.}\label{fig1}
\end{figure}
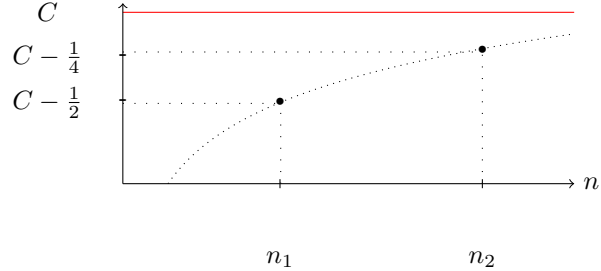

Let $\epsilon >0$, $\epsilon\in\Q\cap(0,1)$ be the admissible decoding error and $\{R_n(\epsilon)\}_{n\in\N}$ be a sequence of achievable rates of capacity achieving codes. It is of interest to determine the conditions under which, for a given $M\in\N$, where $M$ describes the precision of the deviation of $C(P,N)$, for a certain blocklength $n_M$, when the following holds:
\begin{equation}\label{eq:B1}
	R_{n_M}(\epsilon)> C(P,N)-\frac{1}{2^M}.
\end{equation}
This is visualized in Figure~\ref{fig1}.

Next we introduce the definition of time complexity of a computable real number.
\begin{defn}
 Let $t$ be an integer function. The \emph{time complexity} of a computable real number $x$ is bounded by $t$ if there exists a Turing machine that computes, on each input $n\in\N$, a dyadic rational number $d$ in $t(n)$ moves such that $|d-x|\leq2^{-n}$.
\end{defn}
Further, we introduce the concept of a polynomial time computable number.
\begin{defn}
A real number $x$ is \emph{polynomial time computable} if its time complexity is bounded by a polynomial function $p$.
\end{defn}
We now follow with the introduction of a polynomial time computable sequence.
\begin{defn}\label{def:pol}
 	Let $\{\alpha_{n}\}_{n\in\N}$ be a computable sequence of computable numbers. This sequence is computable in polynomial time if there exists a polynomial $p\colon N\times\N\rightarrow \N$, such that for all $n \in \N$ and for each $M \in \N$, a number $\alpha_{n,M} \in \Q$ is computed in at most $p(n,M)$ steps such that 
 	\begin{equation*}
 		|\alpha_n-\alpha_{n,M}|\leq \frac{1}{2^M}
 	\end{equation*}
 	 holds.
\end{defn}

\begin{rem}
    It holds that $R_n(\epsilon)=\frac{1}{n}\log_2\sM_n(\epsilon)$ for $n\in\N$. Hence, for every $n\in\N$ the achievable rate $R_n(\epsilon)$ is a polynomial time computable number.
\end{rem}
The following theorem is discussed in the rest of the abstract.
\begin{thm}\label{thm1}
    Let $B$ be a polynomial time computable number representing the bandwidth. There exists a strictly positive and polynomial time computable noise power spectrum $N_*$ such that for all sufficient large rational power constraint $P_*$ and for all rational $\epsilon > 0$, the computation of achievable rate sequence $\{R_{n_M}(\epsilon)\}_{M \in \mathbb{N}}$ fulfilling: 
	\begin{equation}\label{eq2}
		R_{n_M}(\epsilon)>C(P_*,N_*)-\frac{1}{2^M}.
	\end{equation}
is in $\#\mathrm{P}_1$.
	
    If $\mathrm{FP}_1 \neq \#\mathrm{P}_1$, then for $N_*$ and for every sufficiently large $P_*\in\Q$, none of the sequences $\{R_{n_M}\}_{{n_M}\in\N}$ satisfying \eqref{eq2} can be computed in polynomial time.
\end{thm}

\begin{rem}
    For every $M\in\N$ we have that $R_{n_M}(\epsilon)$ is a polynomial time computable number. The complexity of computing the sequence $\{R_{n_M}(\epsilon)\}_{n_M\in\N}$ grows faster than any polynomial as $M$ increases.
\end{rem}

\begin{rem}
    Consequently, determining the blocklengths $\{n_M\}_{M\in\N}$, that satisfy \eqref{eq:B1} is not feasible in polynomial time for an ACGN channel with noise power spectral density $N_*$.
\end{rem}

\begin{rem}
   We show that either the sequence of achievable rates $\{R_{n_M}(\epsilon)\}_{n_M\in\N}$ as a function of the blocklength is not a polynomial time computable sequence, or the sequence of blocklength $\{n_M\}_{M\in\N}$ corresponding to the achievable rates with guaranteed distance to capacity is not a polynomial time computable sequence, see \cite{boche2024characterizationj}. 
\end{rem}

\begin{rem}
    Note that Theorem \ref{thm1} is valid for any computable sequence of achievable rates satisfying the relation \eqref{eq:B1}. In theoretical computer science, a distinction is made between computable and non-computable solutions. For strictly positive, computable continuous spectral densities $N$ and computable $P>0$, $C(P,N)$ is always a computable number. However, in computer science, there is a further distinction between feasible and unfeasible problems within the realm of computable solutions.
The feasibility thesis states that a natural problem has a feasible algorithm if and only if the problem has a polynomial time algorithm; see \cite[p.~90]{carlson2006millennium} and \cite{Cook1990ComputationalCO}. Therefore, if $\mathrm{FP}_1 \neq \#\mathrm{P}_1$, then the problem of computing achievable rates under the performance constraint \eqref{eq:B1} is not algorithmically feasible, even for very easily computable performance parameters of the communication system, i.e., $N$ computable in polynomial time and $P \in \Q$.
\end{rem}

Finding important performance metrics for communication systems is an important task in information and communication theory. Computer-assisted search and optimization play a crucial role here. Important questions, such as the computation of the optimal input distribution for discrete memoryless channels or code constructions, cannot be solved algorithmically on Turing machines depending on the communication parameters; see \cite{lee2023optimizer,boche2022turing, boche2022algorithmic} . It is interesting to see that even for simple communication systems, such as point-to-point ACGN channels, the complexity of important performance metrics is very high under typical complexity assumptions, even for fixed and easily computable communication parameters.

\bibliographystyle{IEEEtran}
\bibliography{IEEEabrv,confs-jrnls,references_coloregaussian}


\includegraphics[width=1\textwidth]{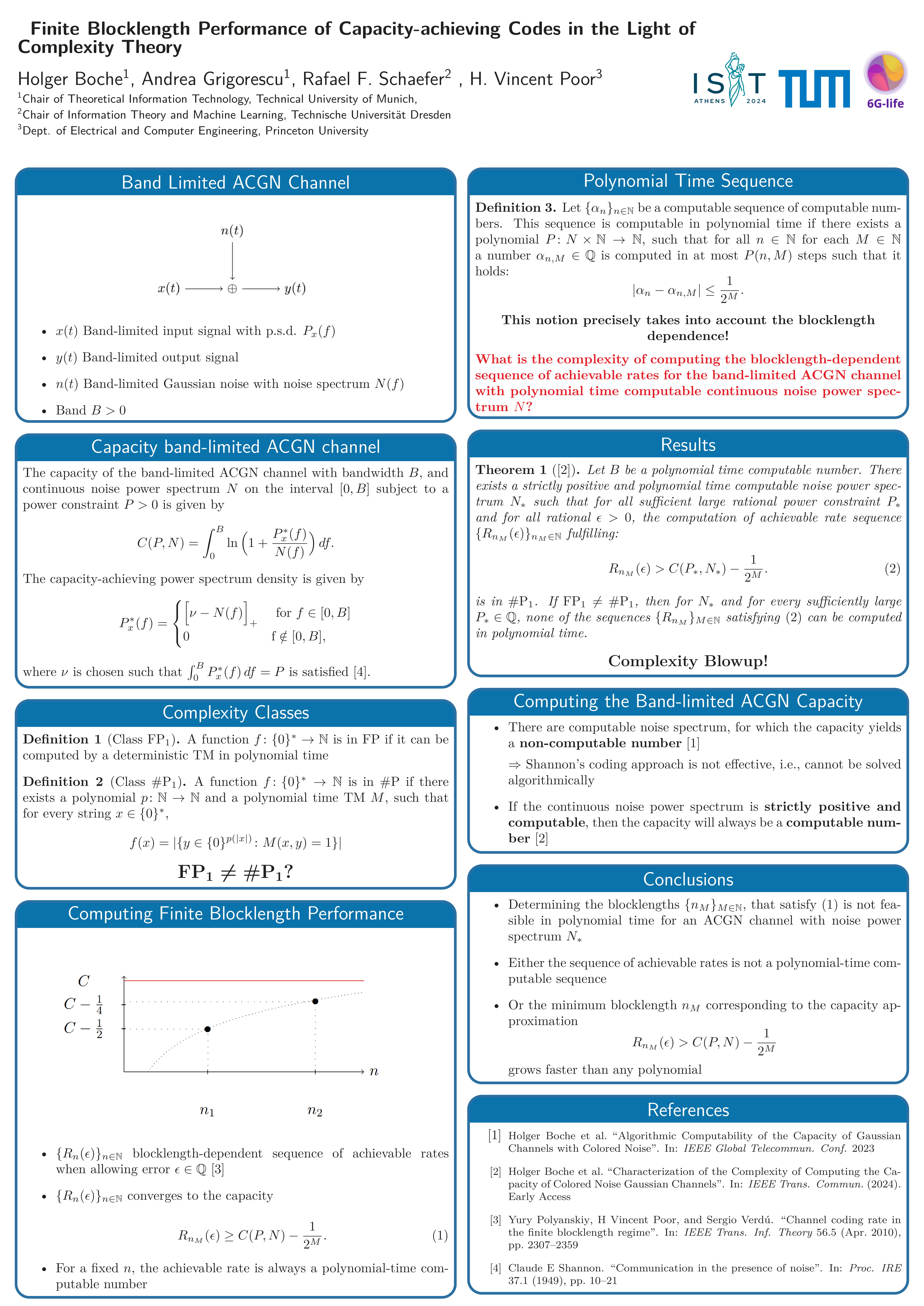}

\end{document}